\documentclass[11pt,twoside]{article}
\usepackage{FUSE2004}
\usepackage{natbib}

\usepackage{epsf}
\usepackage{psfig}
\usepackage{lscape}

\begin{document}
\title{On the Evolutionary Status of Extremely Hot Helium Stars --- 
  are O(He) Stars Successors of RCrB Stars?}
\author{T. Rauch}
\affil{Dr.-Remeis-Sternwarte, 96049 Bamberg, Germany\\
       Institut f\"ur Astronomie und Astrophysik, 72076 T\"ubingen, Germany}
\author{E. Reiff, K. Werner}
\affil{Institut f\"ur Astronomie und Astrophysik, 72076 T\"ubingen, Germany}  
\author{F. Herwig}
\affil{Los Alamos National Laboratory, NM, U.S.A.}  
\author{L. Koesterke}
\affil{Goddard Space Flight Center, Greenbelt, MD, U.S.A.}  
\author{J.W. Kruk}
\affil{Johns Hopkins University, Baltimore, MD, U.S.A.}

\begin{abstract}
95\% of all stars end their lives as white dwarfs. About 20\% of the
hot post-AGB stars are hydrogen deficient. Most of these are the result of a
late helium-shell flash, but the evolutionary status of a fraction of about
10--20\% of the hottest hydrogen-deficient stars, namely four O(He) stars, is
as yet unexplained. They could be the long-searched hot successors of RCrB
stars, which have not been identified up to now. If this turns out to be true,
then a third post-AGB evolutionary sequence is revealed, which is probably the
result of a double degenerate merging process. More generally, understanding
details of merging double degenerate stars is of interest in the context of
SN\,Ia events and hence cosmology.
\end{abstract}

\section{Introduction}
Quantitative spectral analyses of hot hydrogen-deficient post-AGB stars
performed by our groups during the last decade have revealed two distinct
evolutionary sequences. Besides the well-known ``usual'' hydrogen-rich sequence
which is established by central stars of planetary nebulae from early post-AGB
until the hot white dwarf stages, a hydrogen-deficient sequence has been
discovered. It is composed of Wolf-Rayet type central stars which evolve
into PG1159-stars and finally might evolve into non-DA white dwarfs. Our analyses reveal
that the atmospheres of the hydrogen-deficient stars are dominated by carbon,
helium, and oxygen (see reviews by Werner 2001, and Koesterke 2001). A
typical abundance pattern is He:C:O = 33:50:17 (by mass), which is for example
found for the prototype PG1159-035. It has long been argued that these stars
are the result of a late helium-shell flash (late thermal pulse, LTP). The
occurrence of such a flash (the re-ignition of helium-shell burning in a
post-AGB star or white dwarf), has been predicted in earlier investigations by
Iben et al\@. (1983). But only recently we were able to quantitatively explain the
observed surface chemistry with evolutionary models (Herwig et al\@. 1999). Flash
induced envelope mixing causes ingestion and burning of hydrogen. At the same
time, the star evolves back onto the AGB (``born-again AGB star'') and retraces
post-AGB evolution for a second time, but now as a hydrogen-deficient star.
Prominent examples for ongoing ``born-again'' events are FG~Sge and Sakurai's
object.

During the course of our studies of hydrogen-deficient post-AGB stars we have
identified a small group of four extremely hot objects which have almost pure
helium line spectra in the optical. These are classified as O(He) stars following
M\'endez (1991). Our analyses indeed find helium dominated
atmospheres with trace amounts of CNO elements, if detectable at all 
(Tab.\,1). While our evolutionary models can explain the rich diversity of
different He/C/O patterns in Wolf-Rayet and PG1159-stars, they never result in
such helium-dominated surface abundances. It is therefore natural to speculate
on the existence of a third post-AGB evolutionary sequence and its origin. Our
extremely hot helium-rich stars could be the long-searched progeny of the RCrB
stars, which are relatively cool ($T_\mathrm{eff}$ around 10\,000\,K) stars with
helium-dominated atmospheres, too.

\begin{table}[ht]
\caption[]{Parameters of the four known O(He) stars, determined by our analyses
of optical spectra (Rauch et al\@. 1998). Typical uncertainties are: 
$T_\mathrm{eff}$ $\raisebox{0.20em}{{\tiny \hspace{0.2mm}\mbox{$\pm$}\hspace{0.2mm}}}$10\,\%, $\log g$
$\raisebox{0.20em}{{\tiny \hspace{0.2mm}\mbox{$\pm$}\hspace{0.2mm}}}$0.5\,dex, abundance ratios 
$\raisebox{0.20em}{{\tiny \hspace{0.2mm}\mbox{$\pm$}\hspace{0.2mm}}}$0.3\,dex. For comparison, the last two
lines give the mean element abundances of the majority RCrB stars and the
peculiar RCrB star V854~Cen, respectively (Rao \& Lambert 1996). The scatter
around the mean C, N, O, and Si abundances is 
$\raisebox{0.20em}{{\tiny \hspace{0.2mm}\mbox{$\pm$}\hspace{0.2mm}}}$0.15,
$\raisebox{0.20em}{{\tiny \hspace{0.2mm}\mbox{$\pm$}\hspace{0.2mm}}}$0.21, 
$\raisebox{0.20em}{{\tiny \hspace{0.2mm}\mbox{$\pm$}\hspace{0.2mm}}}$0.46,
$\raisebox{0.20em}{{\tiny \hspace{0.2mm}\mbox{$\pm$}\hspace{0.2mm}}}$0.18\,dex, respectively.}\vspace{-2mm}
\label{php}
\begin{center}
\begin{tabular}{lcr@{.}lr@{.}lr@{.}lr@{.}lr@{.}l}
\hline\noalign{\smallskip}
\multicolumn{1}{c}{} &
\multicolumn{1}{c}{$T_\mathrm{eff}$} &
\multicolumn{2}{c}{$\log g$} &
\multicolumn{2}{c}{H\,/\,He} &
\multicolumn{2}{c}{C\,/\,He} &
\multicolumn{2}{c}{N\,/\,He} &
\multicolumn{2}{c}{O\,/\,He}  \\
\cline{5-12}
\noalign{\smallskip}
\multicolumn{1}{c}{} &
\multicolumn{1}{c}{kK} &
\multicolumn{2}{c}{cgs} &
\multicolumn{8}{c}{number ratio} \\
\noalign{\smallskip}
\hline
\noalign{\smallskip}
K\,1-27  & 105 & ~6&5 & \hbox{}\hspace{2mm}\raisebox{0.20em}{{\tiny \mbox{$<$}}} 0&2 & \raisebox{0.20em}{{\tiny \mbox{$<$}}} 0&005 & 0&005 & \multicolumn{2}{c}{} \\
LoTr\,4  & 120 &  5&5 &      0&5 & \raisebox{0.20em}{{\tiny \mbox{$<$}}} 0&004 & 0&001 & \raisebox{0.20em}{{\tiny \mbox{$<$}}} 0&008       \\
HS\,1522+6615 & 140 &  5&5 &      0&1 &      0&003 & \multicolumn{4}{c}{}           \\
HS\,2209+8229 & 100 &  6&0 & \hspace{2mm}\raisebox{0.20em}{{\tiny \mbox{$<$}}} 0&2 & \multicolumn{6}{c}{} \\
\noalign{\smallskip}
\hline
\noalign{\smallskip}
majority RCrB &          &   \multicolumn{2}{c}{}  & \hspace{2mm}\raisebox{0.20em}{{\tiny \mbox{$<$}}} 0&0001&    0&010 & 0&004 & 0&005 \\
V854 Cen      &          &   \multicolumn{2}{c}{}  &      0&5   &    0&030 & 0&0003& 0&003\\
\noalign{\smallskip}
\hline
\end{tabular}
\end{center}
\end{table}

For RCrB stars two origins for their hydrogen-deficiency are discussed in the
literature since many years. Either they are the result of a late thermal pulse
or of a merging process of two white dwarfs (double degenerate scenario). Since
we now can rule out the LTP scenario for helium-dominated post-AGB stars, it
seems inevitable that both, the four hot helium-rich stars and the RCrB stars
are mergers and form another evolutionary post-AGB sequence. We therefore
expect similar heavy element abundances. In order to investigate this possible
evolutionary link it is necessary to compare CNO abundances of both groups,
because detailed theoretical predictions for abundances resulting from a merger
event are not available. The element abundances in RCrB stars are well studied
(Tab.\,1). 

\section{Analysis of FUSE spectra and conclusions}

FUSE spectra of all O(He) stars have been taken during Cycle 2. These have been
analyzed using wind models.  No significant wind features are detected neither in the observation nor
in the model spectra (Fig.\,1). Thus, the mass-loss rates of O(He) stars are not higher
than predicted by radiative-driven wind theory and a change of the surface composition
due to the stellar wind is unlikely. Unfortunately, the FUSE spectra do not show isolated
metal lines and thus, allow to determine upper limits for abundances only (ongoing
study).

\begin{figure}[ht]
\plotone{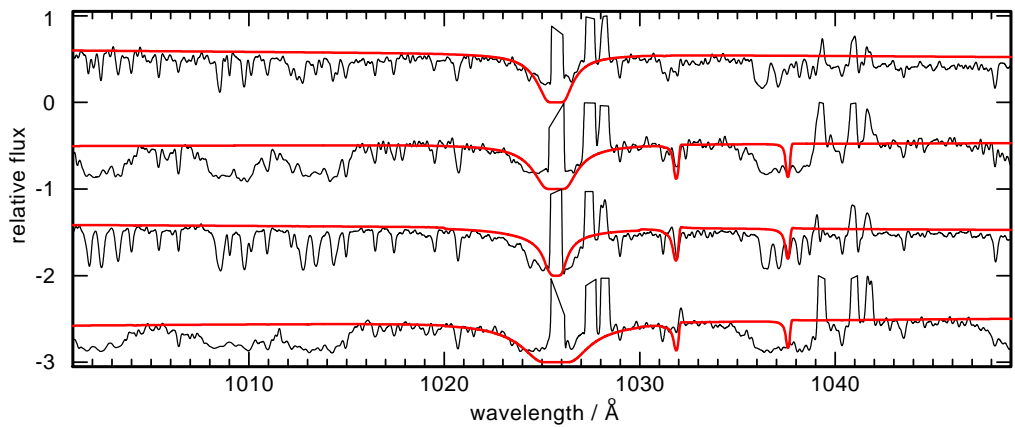}
\caption{Section of the FUSE spectra (around the O\,{\sc vi} $\lambda\lambda$
1031.9, 1037.6\,\AA\ resonance doublet)
of HS\,1522+6615, the central stars of LoTr\,4 and K\,1-27, and HS\,2209+8229 
(top to bottom) compared with synthetic spectra which consider mass loss as predicted
by Pauldrach et al\@. (1988) with mass-loss rates of $\log \dot{M} / \mathrm{M_\odot/yr} = -7.6, -7.7, -9.1, -9.7$,
respectively. The emission features found in the spectra are due to airglow, and the absorption features not found 
in the models are interstellar, predominantly H$_2$
}
\end{figure}

\acknowledgements{T.R\@. is supported by the DLR under grant 50\,OR\,0201.}

\end{document}